\documentclass[oneside]{amsart}
\usepackage[utf8]{inputenc}
\usepackage[letterpaper,margin=1in]{geometry}
\usepackage{amsthm}
\usepackage{amsfonts}
\usepackage{graphicx}
\usepackage{amsmath}
\usepackage{amssymb}
\usepackage{array}
\usepackage{multicol}
\usepackage{multirow}
\usepackage{mathtools}
\usepackage{hyperref}
\usepackage{hhline}
\usepackage{changepage}
\usepackage[english]{babel}
\usepackage[dvipsnames]{xcolor}
\usepackage{appendix}
\usepackage{pdflscape}
\usepackage[figureright]{rotating}
\usepackage[ruled,vlined]{algorithm2e}
\usepackage{float}
\usepackage{subcaption}
\usepackage[foot]{amsaddr}

\hypersetup{
    colorlinks=true,
    linkcolor=blue,
    filecolor=magenta,
    urlcolor=blue,
    citecolor=blue
}
\makeatletter
\newtheorem*{rep@theorem}{\rep@title}
\newcommand{\newreptheorem}[2]{%
\newenvironment{rep#1}[1]{%
 \def\rep@title{#2 \ref{##1}}%
 \begin{rep@theorem}}%
 {\end{rep@theorem}}}
\makeatother

\newenvironment{amssidewaysfigure}
  {\begin{sidewaysfigure}\vspace*{.73\textwidth}\begin{minipage}{\textheight}\centering}
  {\end{minipage}\end{sidewaysfigure}}

\newtheorem{theorem}{Theorem}
\newtheorem{corollary}[theorem]{Corollary}
\newreptheorem{corollary}{Corollary}
\newtheorem{definition}{Definition}
\newreptheorem{theorem}{Theorem}

\newreptheorem{lemma}{Lemma}

\DeclarePairedDelimiter{\ceil}{\lceil}{\rceil}
\DeclarePairedDelimiter{\floor}{\lfloor}{\rfloor}

\title{Optimal Legislative County Clustering in North Carolina}
\author{Daniel Carter$^\text{\MakeLowercase{a}}$}
\author{Zach Hunter$^\text{\MakeLowercase{a}}$}
\author{Dan Teague$^\text{\MakeLowercase{a}}$}
\address{$^a$North Carolina School of Science and Mathematics, Durham NC}

\author{Gregory Herschlag$^\text{\MakeLowercase{b}}$}
\address{$^b$Department of Mathematics, Duke University, Durham NC}
\author{Jonathan Mattingly$^{\text{\MakeLowercase{b}},\text{\MakeLowercase{c}}}$}
\address{$^c$Department of Statistical Science, Duke University, Durham NC}
\date{}

\begin{document}

\maketitle


North Carolina's constitution requires that state legislative districts should not split counties. However, counties must be split to comply with the ``one person, one vote'' mandate of the U.S. Supreme Court. Given that counties must be split, the North Carolina legislature and courts \cite{lakeStephensonBartlett2002} have provided guidelines that seek to reduce counties split across districts while also complying with the ``one person, one vote'' criteria. Under these guidelines, the counties are separated into clusters; each cluster holds a number of districts based on its population. Districts may not span clusters, meaning that each cluster forms an independent set of districts in the sense that it can be subdivided into districts without affecting other clusters. In many county clusters, there are more than one district.

The guidelines for clustering counties were clarified by the courts in 2015 \cite{DicksonRucho2015}. In 2017 the districting plans drawn in 2011 for the North Carolina House and Senate were found to be racially gerrymandered.  The remedy accounted for the courts' 2015 clustering clarification and redrew some of the clusters and districts for use in the 2018 elections. The enacted set of clusters in both the state House and state Senate were reported to be optimal in that the remedy produced the largest number of county clusters possible while following the outlined procedure. However, no transparent validation of this claim was provided in the public domain.

The primary goal of this work is to develop, present, and publicly release an algorithm to optimally cluster counties according to the guidelines set by the court in 2015. We use this tool to investigate the optimality and uniqueness of the enacted clusters under the 2017 redistricting process. We verify that the enacted clusters are optimal, but find other optimal choices. We emphasize that the tool we provide lists \textit{all} possible optimal county clusterings.

We also explore the stability of clustering under changing statewide populations and project what the county clusters may look like in the next redistricting cycle beginning in 2020/2021. In studying the stability of these clusters, we find that their structure may be highly susceptible to small fluctuations in the population; on average approximately one third of the clusters change each year. In addition, we compare the existing guidelines with an alternative interpretation of how counties might be minimally split. As part of this report, we provide code,\footnote{Git repository available at https://git.math.duke.edu/gitlab/gjh/countycluster.git} along with documentation and examples, which may be used by the public to independently verify the North Carolina legislative district clusters during the next redistricting cycle.

The report is organized as follows. In Section~\ref{sec:LegislativeDistricting}, we lay out the general redistricting problem for North Carolina in light of various laws and legal precedents. 
We then explain how this leads to the ``county clustering'' problem and the court-sanctioned procedure for resolving it. In Section~\ref{sec:Algorithm}, we describe our algorithm and give the needed theoretical justification to ensure it produces all of the optimal county clusters. In Section~\ref{sec:2017Clusters}, we apply our algorithm to the 2010 census data which was used to construct the redistricting used in the 2017 and 2018 elections. We show that the actual clustering used was optimal, but that they were not the only choices. 
In Section~\ref{sec:StabilityOverTime}, we study the stability of the optimal county clusters over time if one were to calculate the optimal clusters each time population estimates are updated by the census. 
In Section~\ref{sec:2020}, we analyze what the clusters might look like in 2020 using current population forecasts. 
Lastly, in Section~\ref{sec:metrics}, we alter the court-mandated optimization procedure and find county clusterings which split fewer counties than the hierarchical procedure. In Section~\ref{sec:Discussion}, we give some concluding observations and summarize our results.

\noindent \textsc{Acknowledgments: } We are thankful to Blake Esselstyn and Eddie Speas bringing this important question to our attention as well as generally educating us about legal and GIS issues involved. This project began as a project in a year long research experience hosted at Duke University for students from the North Carolina School of Science and Mathematics. The authors acknowledge suport from a grant from \textit{Imagine North Carolina First} which partially supported GH and the summer work of DC and ZH as well as the Mathematics Department and the Rhodes Information initiative at Duke for hosting this work.
\section{Legislative Districting in North Carolina}
\label{sec:LegislativeDistricting}


North Carolina elects its general assembly consisting of 50 senators and 120 representatives by districts. Article 2 of North Carolina's constitution states \cite{NorthCarolinaConstitution} that

\begin{enumerate}
    \item Each [legislator] shall represent, as nearly as may be, an equal number of inhabitants, the number of inhabitants that each [legislator] represents being determined for this purpose by dividing the population of the district that he represents by the number of [legislators] apportioned to that district;
    \item Each [legislative] district shall at all times consist of contiguous territory;
    \item No county shall be divided in the formation of a [legislative] district;
    \item When established, the [legislative] districts and the apportionment of [legislators] shall remain unaltered until the return of another decennial census of population taken by order of Congress.
\end{enumerate}

Additionally, districts must comply with the federal Voting Rights Act (VRA), which mandates the construction of several districts where minority votes have a reasonable chance to effect the election outcome. 
This translates into requiring some districts with higher concentrations of minority votes.  
For the purposes of this paper, it is assumed that it is always possible to comply with the VRA once the clusters are set, so we will ignore this constraint.

Constraint (1) was clarified in 2002 in the case \textit{Stephenson v. Bartlett} \cite{lakeStephensonBartlett2002}, when North Carolina's Supreme Court ruled that all districts must be single-member districts, and that the population of each district must be within 5\% of the ideal population, which is $1/50$th of the state population for Senate districts and $1/120$th of the state population for representative districts.

The definition of ``contiguous'' in constraint (2) was clarified in 2003 in \textit{Stephenson II} \cite{lakeStephensonBartlettII2003}. In North Carolina, districts which are contiguous are not allowed to be connected by just one point (e.g. if four counties meet at a corner, a district cannot go across the corner to connect opposite counties). However, districts \textit{are} considered contiguous even if they are only connected by water.

Constraint (3) is known as the ``Whole County Provision,'' or WCP \cite{lakeStephensonBartlett2002}. Due to the population distribution of North Carolina, it is impossible to satisfy constraint (1) while keeping every county whole. For example, several counties contain significantly more than $1/120$th of the state population, so they must be split into multiple House districts. Counties may also have too much population to be one district but too little to be split into two districts. In this case, a district must be drawn which crosses county lines.

The process chosen to comply with both of these constraints is to first group counties together into county clusters, then draw districts that do not cross cluster lines. 
The more clusters formed and the fewer counties in each cluster, the fewer districts must be split between counties.\footnote{See Section~\ref{sec:metrics} and Appendix~\ref{app:metrics} for a more thorough discussion of this fact.} North Carolina's Supreme Court ruled that a legal clustering is one that contains the most single-county clusters, then among those options, one that contains the most two-county clusters, and so on \cite{DicksonRucho2015}. As mentioned earlier, we assume that it is always possible to construct VRA districts regardless of the clustering chosen, so we have only considered NC Constitution constraints (1-3) in this paper.

\section{Mathematical Overview}
\label{sec:Algorithm}

The legal definition that concluded the previous section may be recast into mathematical terms in the following way
\begin{definition}
Consider two county clusterings $A$ and $B$. 
If clustering $A$ has $a_1$ single-county clusters, $a_2$ two-county clusters, etc. and clustering $B$ has $b_1$ single-county clusters, $b_2$ two-county clusters, etc., clustering $A$ is \textit{preferred} over $B$ if $a_n>b_n$ for some $n$ and $a_m=b_m$ for all $m<n$. 
A clustering is a \textit{legal county clustering} (equivalently \textit{optimal county clustering}) if no clustering is preferred over it.
\label{def:optClust}
\end{definition}

We develop an algorithm to determine all optimal county clusterings.  
At the highest level, the basic algorithm can be described as follows: 
First, identify which single counties can be divided into an integer number of districts, with each district being within 5\% of the ideal district population dictated by the ``one person, one vote'' principle, and such that the remaining counties can still be grouped into legal clusters. 
Then among the remaining counties, we look for all pairs of contiguous counties which contain an integer number of districts within 5\% of the ideal population, where remaining counties can still be grouped into legal clusters. 
This process is then performed with groups of three contiguous counties, then four, and so on. 
Eventually all of the counties are placed in a county cluster, with each cluster having an assigned number of districts it should be subdivided into. Pseudocode for the algorithm is presented in Algorithm~\ref{alg:highlevel}.  

\begin{algorithm}
    \KwIn{Populations, adjacencies, number of districts.}
    \KwOut{Optimal clustering.}
    Let all counties be unassigned to a cluster\;
    Let $n=1$\;
    \While{any county is unassigned}{
        Let $S$ be the largest disjoint set of $n$-county clusters made from unassigned counties, such that unassigned counties not in $S$ can still be made into legal clusters $(*)$\;
        Assign counties according to $S$\;
        Increment $n$\;
    }
    \Return{the county cluster assignment}\;
    \caption{High-level pseudocode for the county clustering algorithm}
    \label{alg:highlevel}
\end{algorithm}

The most difficult part this procedure is verifying that the remaining counties might still form a collection of legal county clusters (marked $(*)$ in Algorithm~\ref{alg:highlevel}).
To find such a set of clusters, we construct a search tree and use a branch-and-bound depth-first search method to traverse it. Each level of the tree represents adding one county cluster consisting of $n$ counties, so the depth of the tree corresponds to the number of clusters with $n$ counties we have added. The leaves at the deepest level of the tree represent all largest sets of clusters containing $n$ counties.
In constructing this tree, we must guarantee that it is still possible to cluster all remaining unassigned counties at each level. The following theorem allows us to quickly characterize whether it is possible to cluster these remaining counties. A more verbose formulation of the theorem is given and proven in Appendix~\ref{app:decision}.

\begin{reptheorem}{thm:decision}[Short Version]

A set of counties $C$ can be clustered into $D$ districts if and only if each contiguous subset of $C$, $c_k$, can be divided into $d_k$ districts subject to the constraint that $\sum d_k = D$, where $D$ and $d_k$ are all positive integers.

\end{reptheorem}

The search tree is far too large to search entirely. We can reduce the search space dramatically using two mathematically-rigorous bounds on ``how good'' a solution under a particular branch could be. These optimizations allow us to search only the parts of the tree which could lead to optimal solutions. The full details behind the algorithm and bounds can be found in Appendices~\ref{app:optimize} and \ref{app:algorithm}.

\section{2017 Clusterings}
\label{sec:2017Clusters}

\begin{figure}
    \centering
    \begin{subfigure}[t]{\textwidth}
        \centering
       \includegraphics{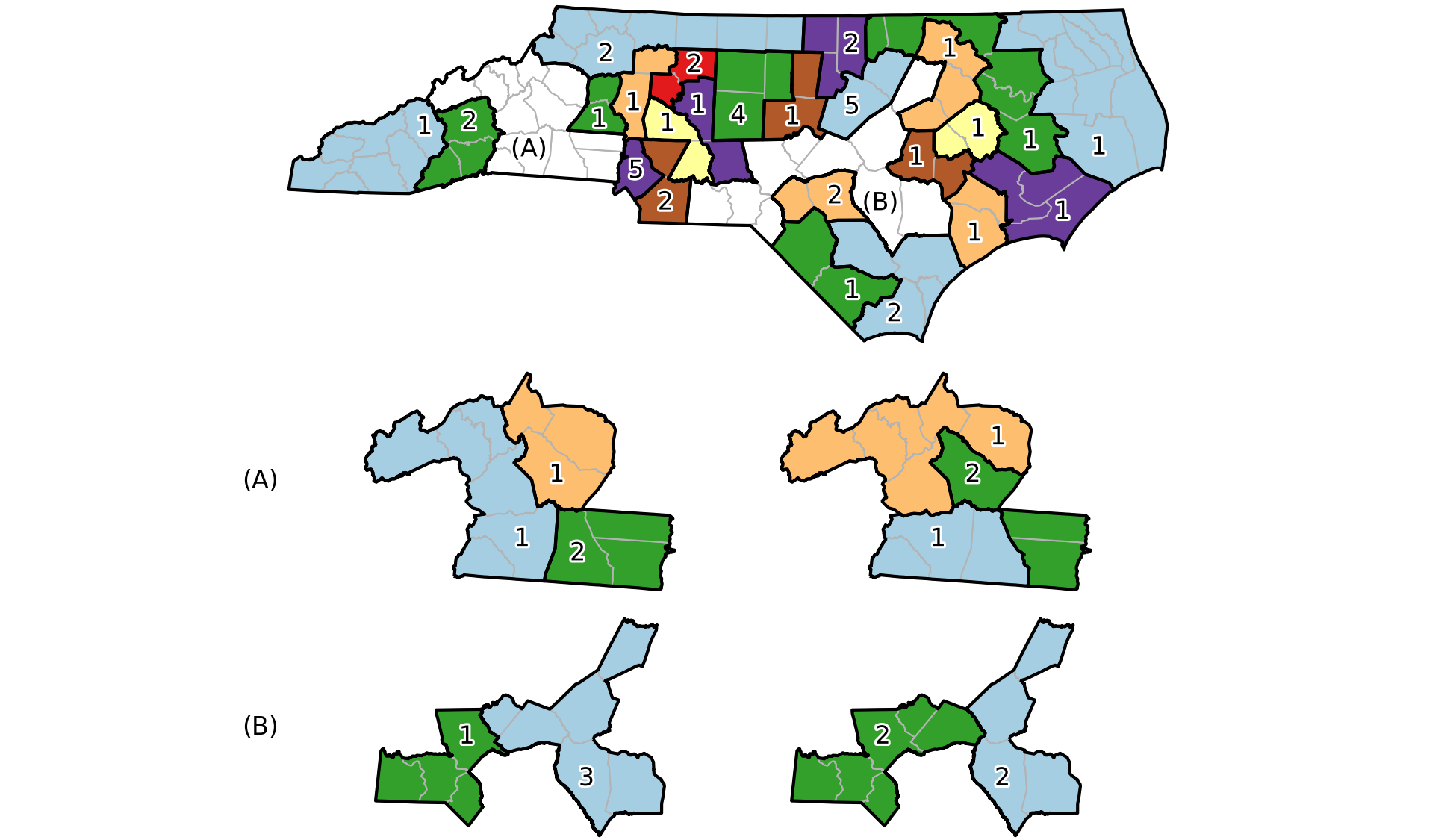}
       \caption{Optimal Senate clusterings}
       \label{fig:Senate2010}
    \end{subfigure}%
    \newline
    \begin{subfigure}[t]{\textwidth}
        \centering
       \includegraphics{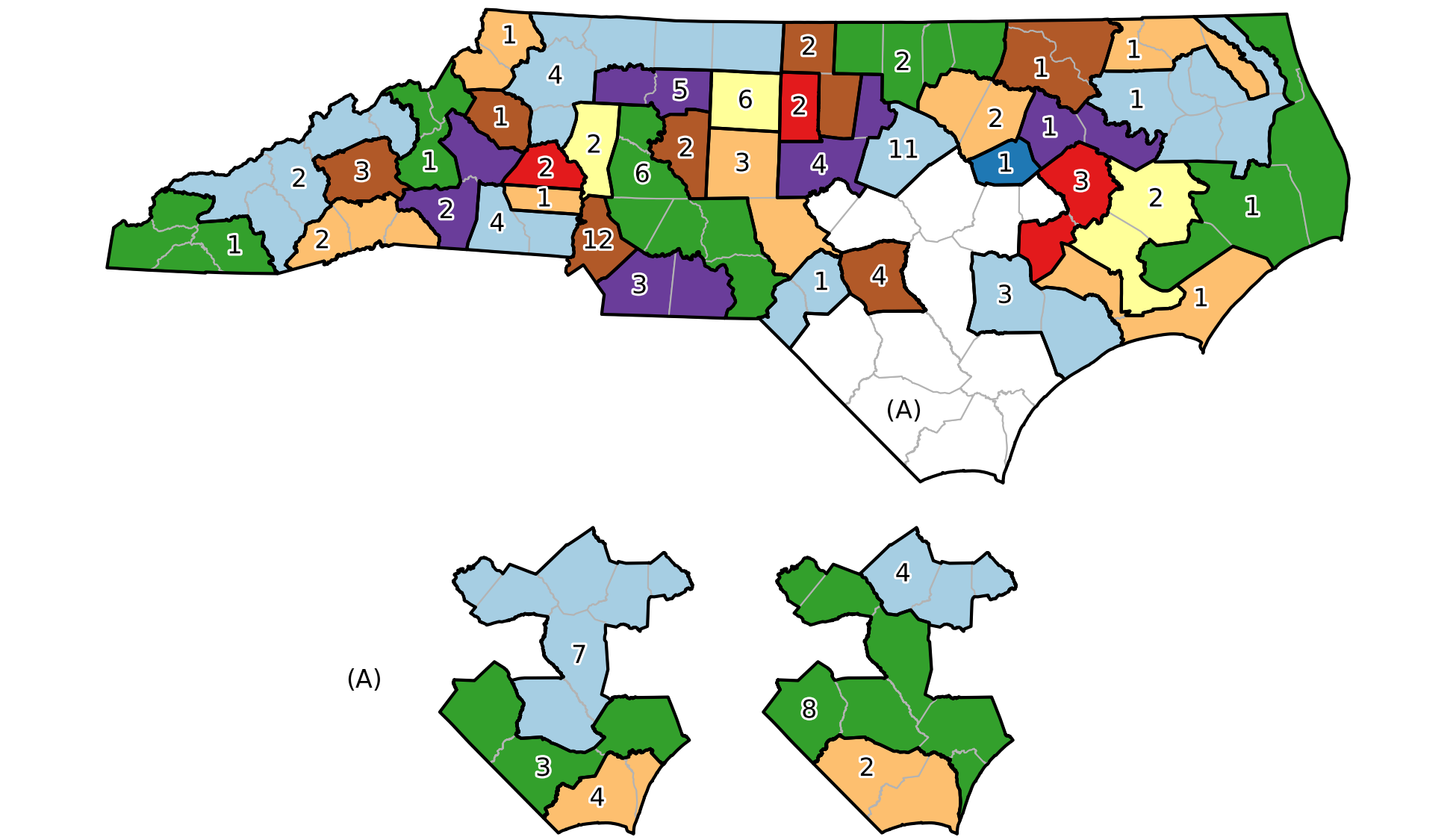}
       \caption{Optimal House clusterings}
        \label{fig:House2010}
    \end{subfigure}
    \caption{The two white regions, labeled (A) or (B) in the state map, have different possible clusters. The map fragments below each state map, also labeled (A) or (B), show the different possible ways to complete the corresponding region of the map. The 2017 clustering, employed in the currently enacted redistricting, used the leftmost option in each set of alternatives.  The colors are for clarity and have no significance.}
    \label{fig:2010}
\end{figure}

Using the clustering algorithm described in Section~\ref{sec:Algorithm}, we determine the optimal county clusterings using the 2010 census data. We find that the clusterings constructed in 2017 and used in the 2018 elections are optimal. However, there are other optimal county clusterings for both the Senate and House districting plans; we find four optimal clusterings in the Senate and two in the House. All possible optimal clusterings are depicted in Figure~\ref{fig:2010} for the Senate and House. In each figure, a map of the entire state demonstrates the clusters (colored) that are found in all optimal maps. Regions which have multiple options are not colored (kept white) in this map and labeled with a letter in parentheses; each option is shown in subsequent maps in the figures. Clusters are labeled with the number of districts they contain. The clusters used in the maps enacted in 2017 are the first option in each region with possible choices.

\begin{figure}
    \centering
    \includegraphics[width=0.75\textwidth]{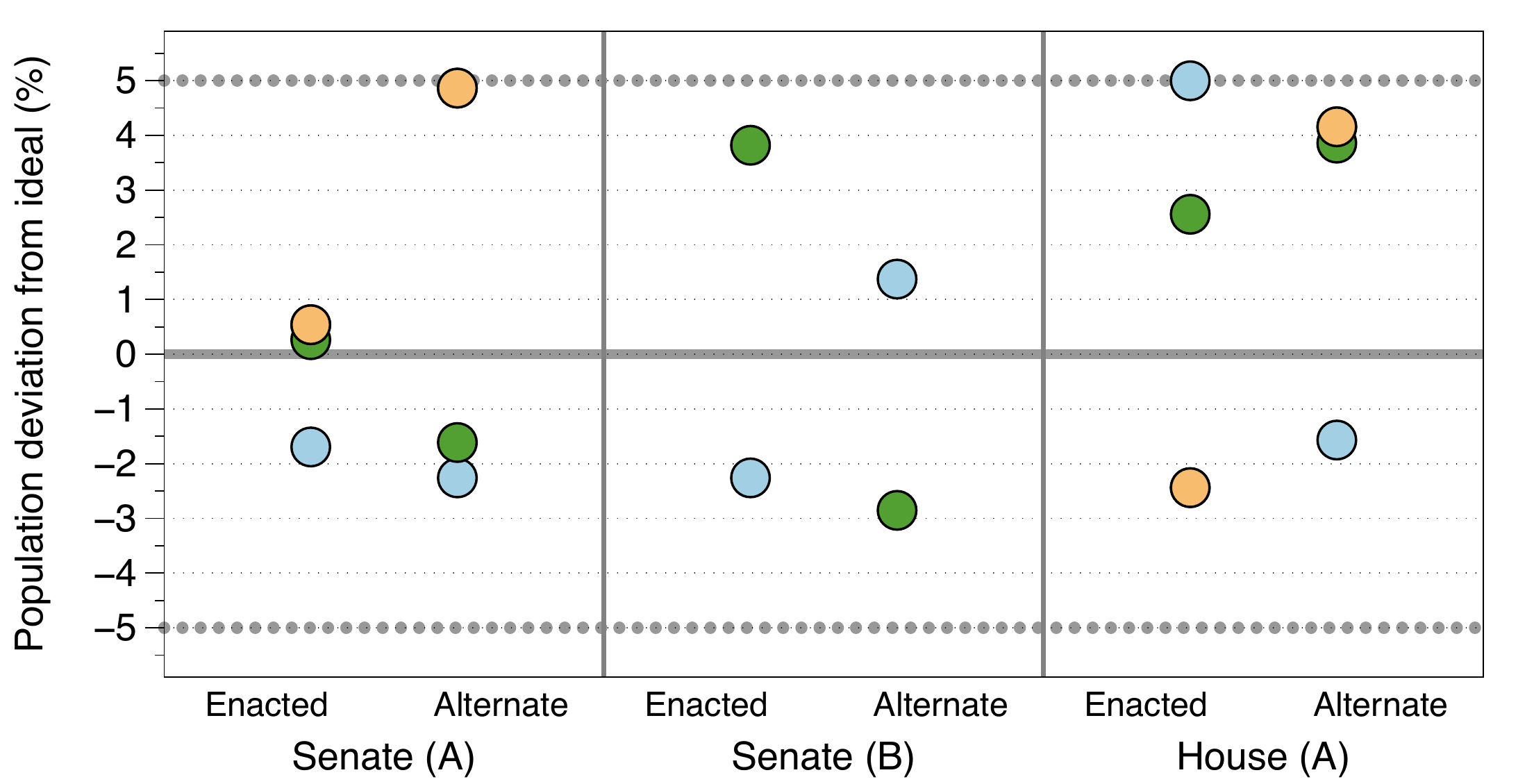}
    \caption{We display the average population deviation for each choice of clusterings using 2010 census data. Each point gives the average percent population deviation from the ideal population if all of the districts within a county cluster are taken to have the same population. The (A) and (B) labels refer to the different regions of choice labeled in Figure~\ref{fig:2010}. 
    The different colors correspond to the color of the county clusters found in Figure~\ref{fig:2010}.
    }
    \label{fig:pop_deviation}
\end{figure}

Although the choices are all equivalent under the definition of optimality, some may be preferable over others in fulfilling the legal requirements for districtings. For example, the deviation from ideal population is not the same between choices. The population deviations in the clusters in each option are summarized in Figure~\ref{fig:pop_deviation}. The numbers shown are the percent difference between each district's population and the ideal population if all districts within a cluster were drawn to have the same population. 

In both the Senate region (B) and House region (A), the alternative, non-enacted, clustering option leads to districts that are closer to the ideal population compared to the enacted clustering. The most striking example is House region (A). The northernmost cluster, which contains 7 districts, had 4.996\% more population than ideal. In fact, one district in the enacted districting slightly exceeds the 5\% threshold \cite{persilySPECIALMASTERRECOMMENDED2017}; this could have been avoided by choosing the other optimal clustering.

\section{Stability of Clusterings Over Time}
\label{sec:StabilityOverTime}

Using the Census Bureau population estimates \cite{AnnualEstimatesResident2019}, we found the optimal clusterings for each year from 2010 to 2018. Note that the Census Bureau estimates the population at July 1 of each year, but the 2010 census represents the population at April 1. Hence, two results are given for 2010 which differ by 3 months.

In Figure~\ref{fig:Senate_years} and Figure~\ref{fig:House_years}, the optimal clusterings for each year are depicted. Clusters which remain the same between adjacent years are colored light gray. Most years had multiple optimal options; the one shown is the one most similar to the previous year.

\begin{amssidewaysfigure}
    \centering
    \includegraphics{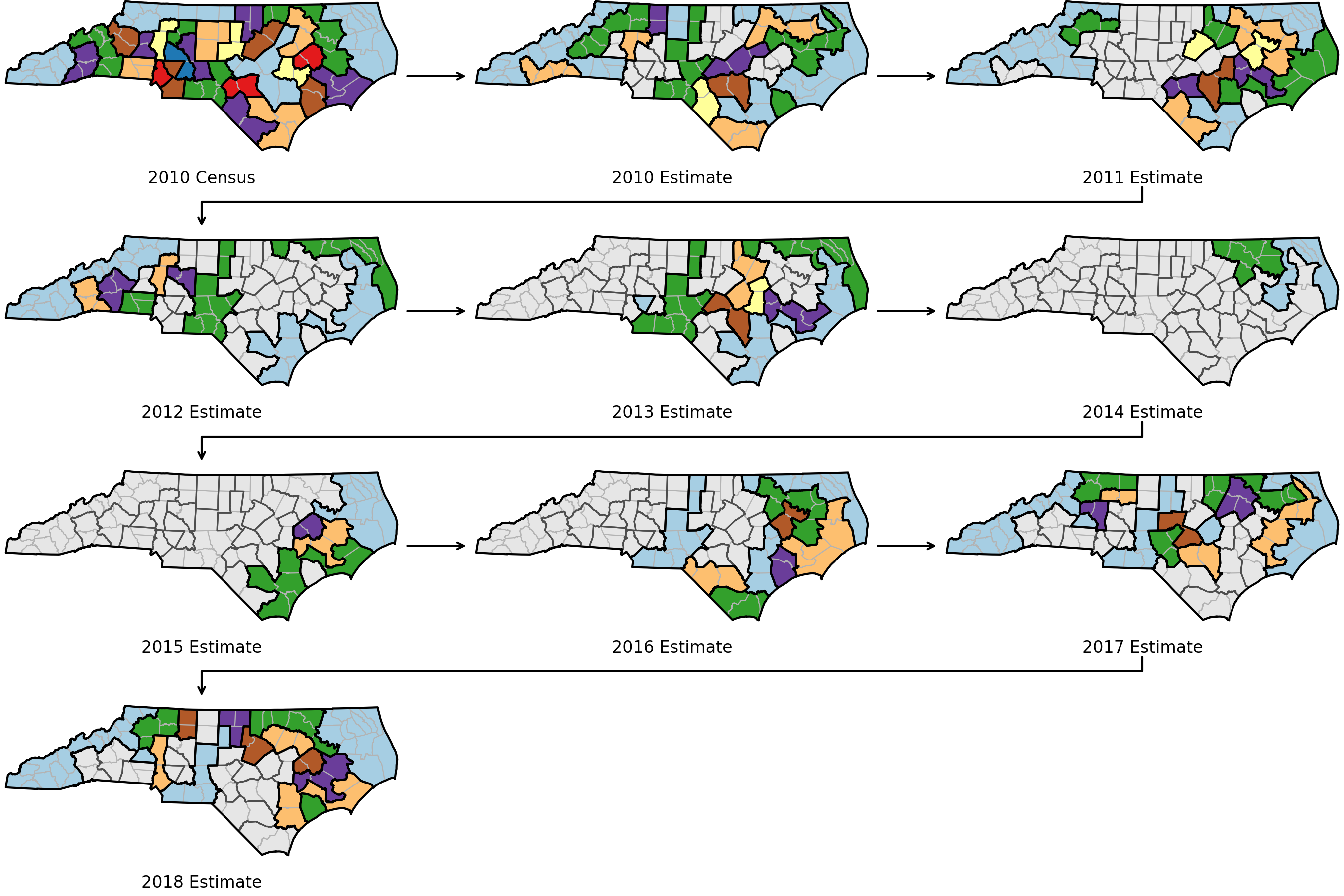}
    \caption{Optimal Senate clusterings from April 2010 to July 2018. The 2010 census reflects the population in April 2010; all estimates reflect the population in July. Grayed out clusters are the as the previous map.  The color choice of each colored cluster is for qualitative purposes only.}
    \label{fig:Senate_years}
\end{amssidewaysfigure}

\begin{amssidewaysfigure}
    \centering
    \includegraphics{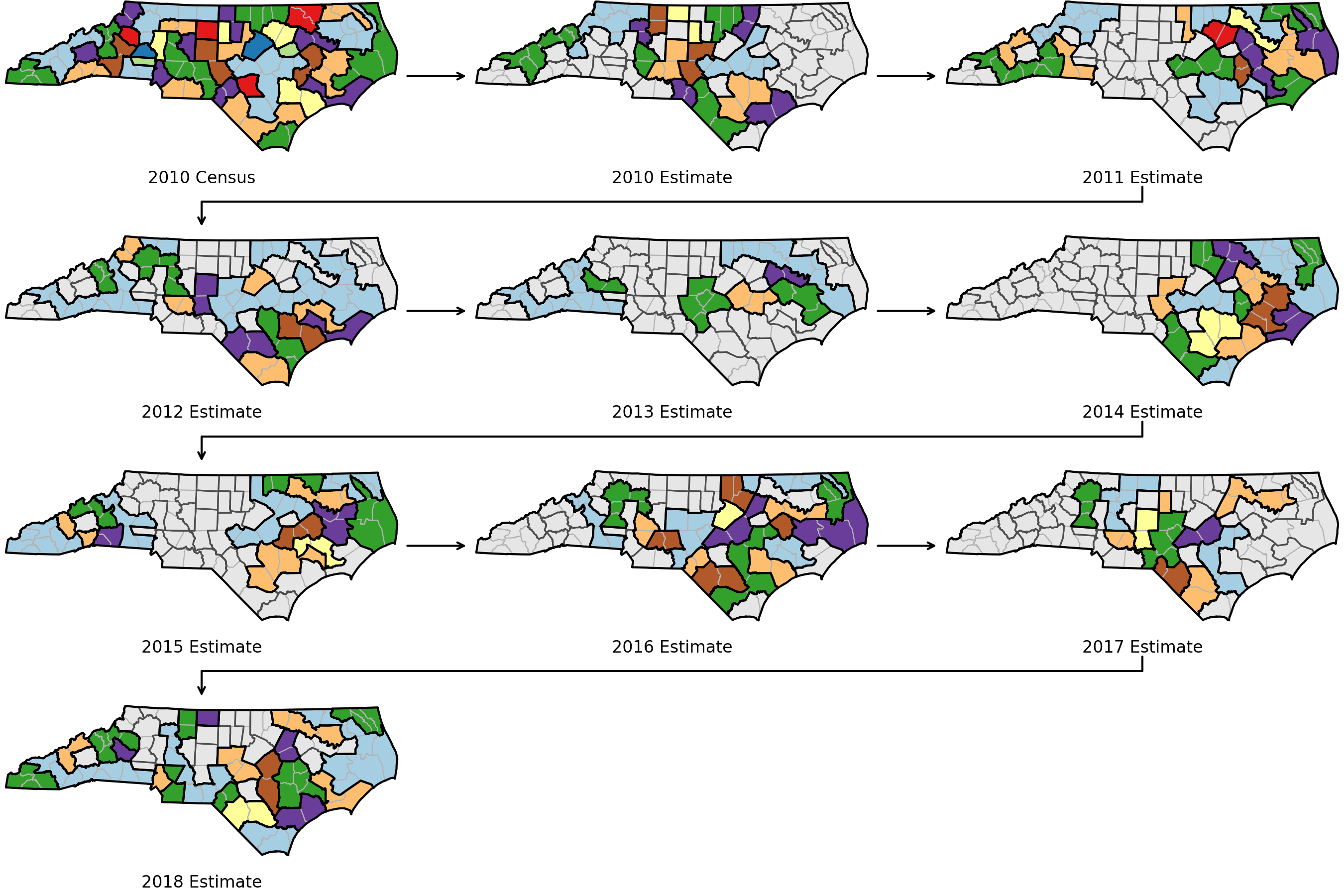}
    \caption{Optimal House clusterings from April 2010 to July 2018. The 2010 census reflects the population in April 2010; all estimates reflect the population in July. Grayed out clusters are the as the previous map.  The color choice of each colored cluster is for qualitative purposes only.}
    \label{fig:House_years}
\end{amssidewaysfigure}

We use several measures to quantify the distance between clusterings. The first measure is the percentage of clusters which are different between the two clusterings (abbreviated DC for ``different clusters''). If the clusterings are the same, the number of different clusters is 0\%, and if every single cluster changed, it is 100\%.

A more sophisticated measure takes into account how much each cluster changes. One way to do this is by the following observation: we choose a county at random and compare its cluster between two clusterings; if those clusters contain many of the same counties, the clusterings are similar. On the other hand, if the clusters have few counties in common, the clusterings are different.

This intuitive notion is made rigorous with \textit{variation of information} (abbreviated VI). The variation of information between two identical clusterings is 0, and a higher number indicates very different clusterings, i.e. a low chance that a randomly chosen county is in a similar-looking cluster. The units of the variation of information are \textit{bits/county}. This measure was previously used in other redistricting papers \cite{fifieldNewAutomatedRedistricting}.\footnote{This measure was used in the early drafts of this paper, but omitted in later drafts.}

The final measure we consider is variation of information divided by the \textit{average population change} (abbreviated VI/APC). The average population change measures the average difference in county population between two data points. Thus, a high VI/APC is achieved when the clustering changed significantly (high variation of information) despite a small change in population (low average population change). This number is a measure of the fragility of clusterings. The units of average population change are \textit{\%/county} so the units of VI/APC are \textit{bits/\%}.

There are generally multiple optimal clusterings in each year. Hypothetically, if the clusterings were redrawn when the new population estimates were released each year, one might desire that the new clustering is the one that preserves the most clusters from the previous clustering. This is how we report the stability metrics: starting with the enacted plans, we update the clusters each time the census bureau updates the population estimate by choosing the clustering which minimizes the number of different clusters; if there are multiple clusterings which have the same number of different clusters, we prefer the ones with a smaller variation of information.

\begin{figure}
    \centering
    \includegraphics[width=0.75\textwidth]{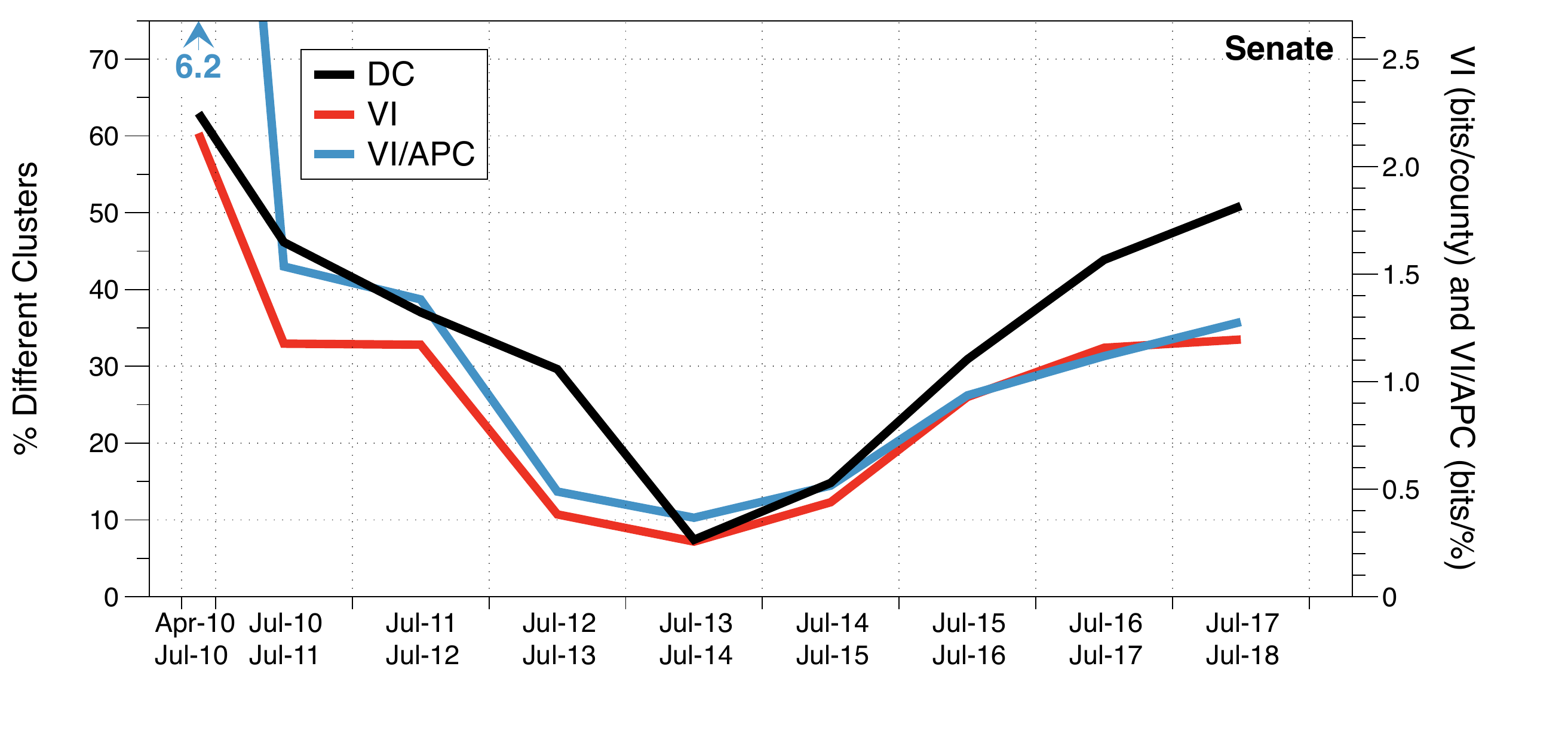}
    \includegraphics[width=0.75\textwidth]{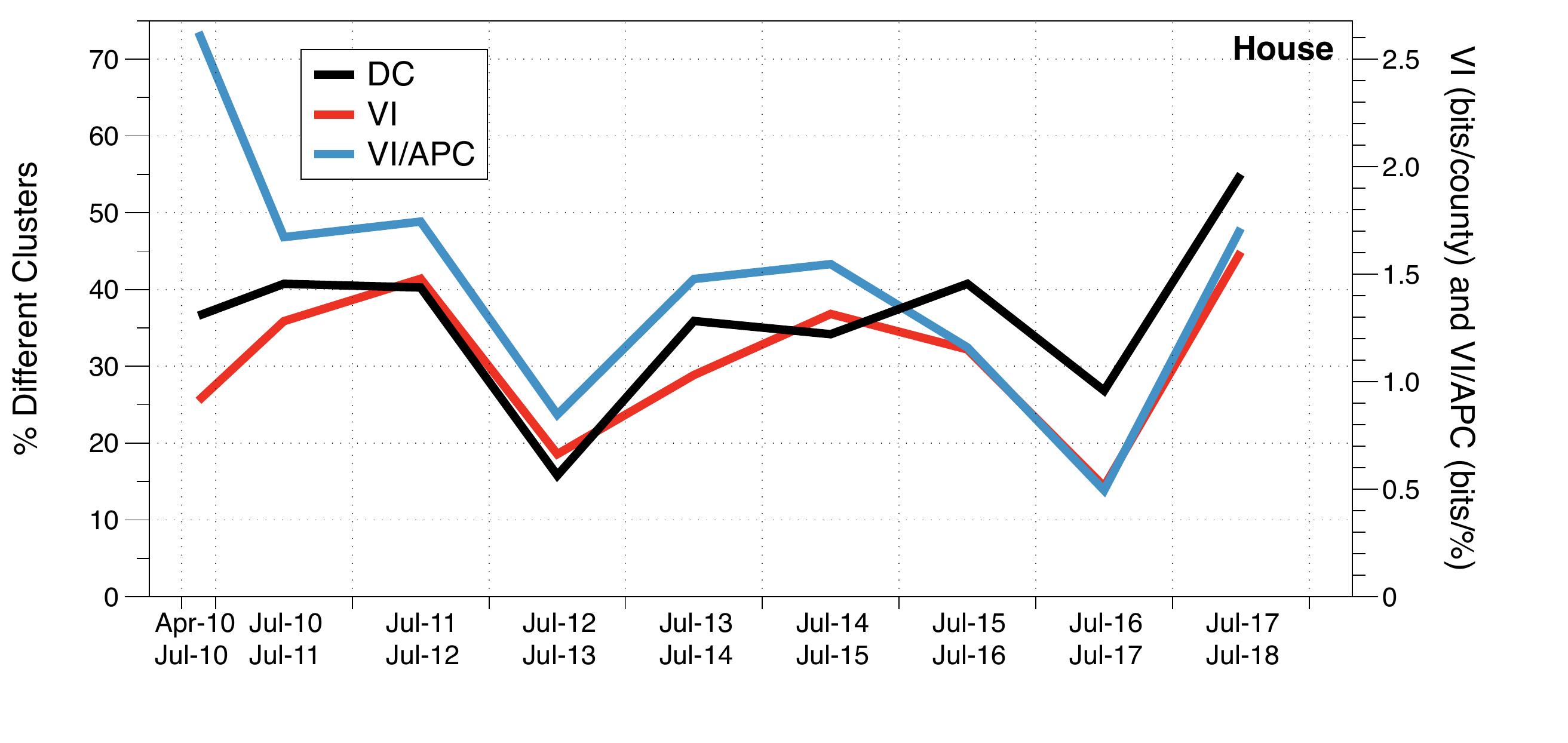}
    \caption{We present three metrics that quantify the minimal amount county clusterings would change across consecutive sets of population data. We find that, on average, roughly a third of the clusters change every time new data is collected (mostly annually, with the exception of a 3 month lapse in 2010); see the black line labeled DC (different clusters). The other two metrics, variation of information (VI) and variation of information by the average population change (VI/APC) are highly correlated with the first. All July dates are census estimates, whereas the April date is the decennial census count.}
    \label{fig:stability}
\end{figure}

The mathematical definitions of different clusters (DC), variation of information (VI), and average population change (APC) can be found in Appendix~\ref{app:stability}. The results from the three metrics are shown in Figure~\ref{fig:stability}.

As can be seen in the figure, the number of different clusters (DC) and the variation of information (VI) are highly correlated. The average population change (APC) of each year was close to 1\%, so the numerical values of VI and VI/APC are close, with the exception of change between the two 2010 data sets.\footnote{The two 2010 data sets represent only 3 months of population change and the APC was just 0.3\%.}
Despite the smaller population difference, the clusters changed significantly in both the House and Senate, causing VI/APC for that data point to be very large. In other words, the clusters constructed using the 2010 census data are very sensitive to population change. Furthermore, on average more than one third of the clusters change each year, meaning that the clusters utilized in any districting cycle may be highly sensitive to timing and quality of the census data.

\section{2020 Projected Clusters}
\label{sec:2020}

\begin{figure*}
    \centering
    \begin{subfigure}[t]{\textwidth}
      \centering
       \includegraphics{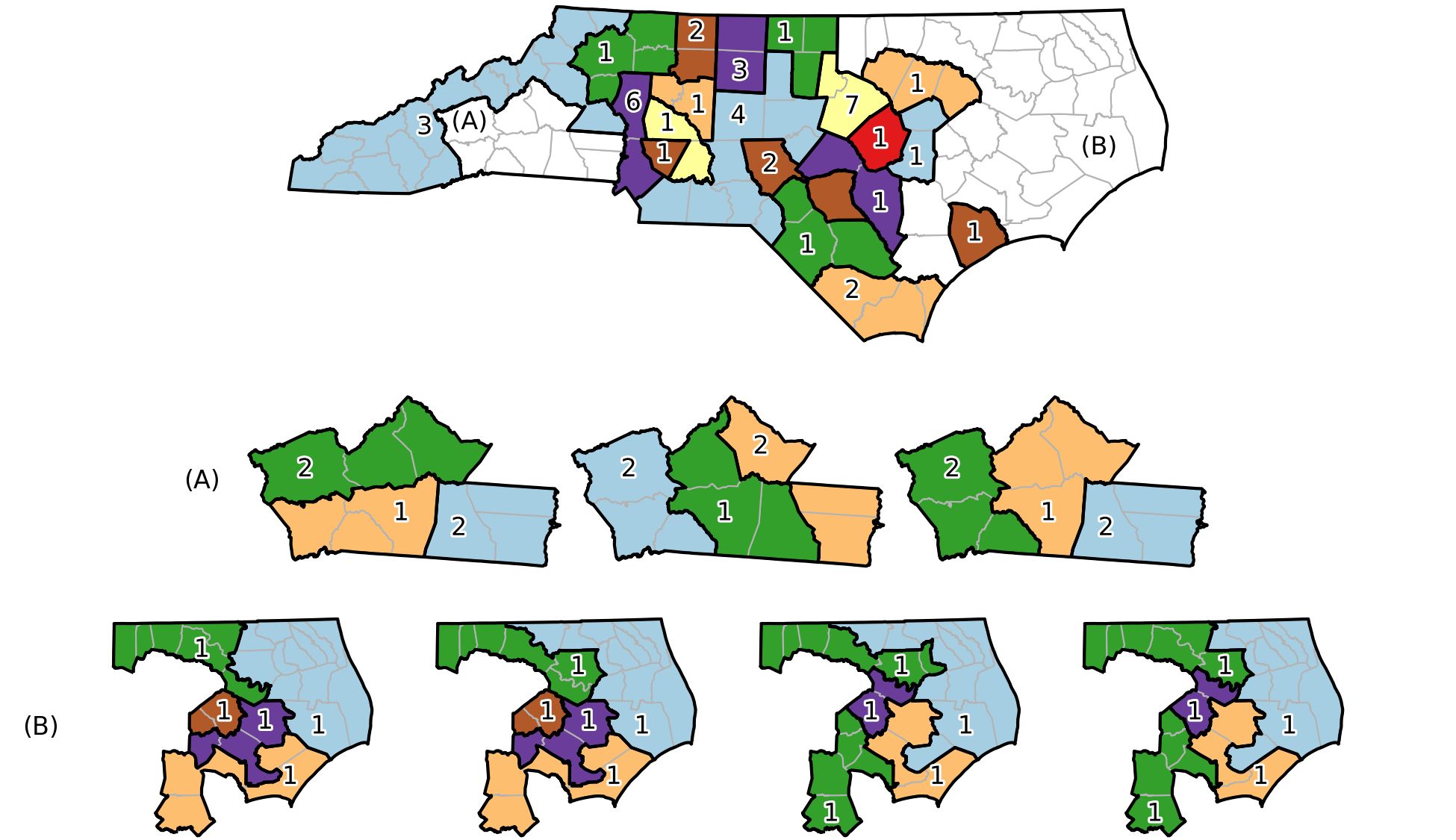}
       \caption{Optimal Senate clusterings using NC OSBM 2020 projections}
       \label{fig:Senate2020}
    \end{subfigure}%
    \newline
    \begin{subfigure}[t]{\textwidth}
        \centering
      \includegraphics{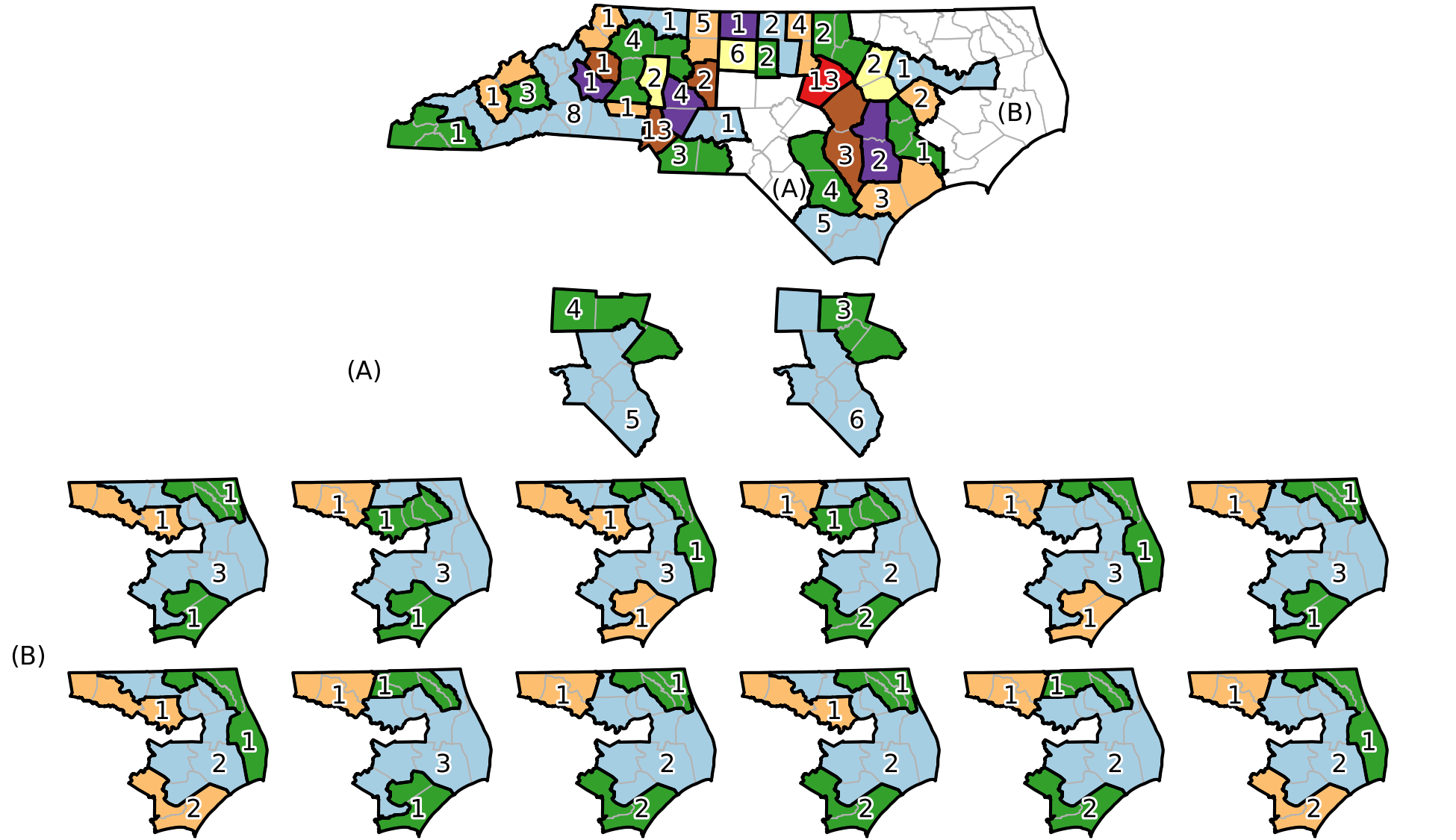}
       \caption{Optimal House clusterings using NC OSBM 2020 projections}
        \label{fig:House2020}
    \end{subfigure}
    \caption{Optimal Senate and House clusterings using NC OSBM 2020 projections. As in Figure~\ref{fig:2010}, the white regions, labeled (A) or (B) in the state map, have different possible clustering. The map fragments below each state map, also labeled (A) or (B), show the different possible ways to complete the corresponding region of the map.}
    \label{fig:2020}
\end{figure*}

Based on the NC OSBM County Population Projections \cite{CountyStatePopulation}, we predict the 2020 clusters. 
The results are shown in Figure~\ref{fig:Senate2020} and Figure~\ref{fig:House2020}. 
This data projects the population in July 2020, and may be far from the actual April 2020 population that the Census will measure due to both uncertainty in the projection and the 3-month time difference. 
As demonstrated before, a small difference in population can still result in a substantial change in the clusters, so it is very unlikely these predictions will be exactly correct, and it is possible they could be very far from what the actual clusters will be.  
We expect the majority of the county cluster to change in the upcoming redistricting cycle.  In the Senate, we find only 2 clusters containing 3 districts that would potentially be preserved in the next redistricting cycle given the current population 
projections.\footnote{The possibly preserved Senate clusters are the Lincoln-Cleveland-Gaston county cluster, which is the southeastern light blue 2-district cluster of the leftmost option (A) in Figure~\ref{fig:Senate2020}, and the Pitt-Greene county cluster which is the brown central 1-district cluster of the leftmost option (B) in Figure~\ref{fig:Senate2020}}
In the House, we find 11 clusters containing 24 districts that would be preserved in the next redistricting cycle;\footnote{The preserved House clusters are the Ashe-Watauga (1 district), 
Caldwell (1 district), 
Caswell-Orange (2 districts), 
Union-Anson (3 districts),
Buncombe (3 districts),
Macon-Cherokee-Clay-Graham (1 district),
Lincoln (1 district),
Davidson (2 districts),
Iredell (2 districts)
Guilford (6 districts) and
Alamance (2 districts)
 clusters; on the map these county clusters are found in Figure~\ref{fig:House2020} to be the orange-yellow northern most district with 1 district to the west, the brown single district cluster immediately south of the previous cluster, the northern-most central light blue two district cluster, the green three district cluster on the southern border, the green three district cluster toward the west of the state, the western-most green one district cluster, orange-yellow one district cluster next to the brown 13 district cluster, the central brown 2 district cluster, the bright yellow 2 district cluster to the west of the previous cluster, the bright-yellow 6 district cluster, and the green two district cluster immediately east of the previous cluster,  respectively.}  Wake and Mecklenburg both are projected to remain single county clusters, however both are projected to increase in the number of districts.  In short, we project that a majority of the state's clusters will be redrawn or have a different number of districts by the next redistricting cycle.



\section{Another Possible Interpretation of the Whole County Provision}
\label{sec:metrics}


The courts could have interpreted the Whole County Provision in ways that would have further reduced the number of county splits and that would have led to different optimal county clusterings.
One idea is that instead of maximizing the number of $n$-county districts at each step, one should instead maximize the \textit{total} number of clusters. As the following theorem shows, this metric is essentially equivalent to minimizing the number of ``county splits,'' where a county with all or parts of $d$ districts counts as $d-1$ county splits. A more precise formulation of this theorem is given and proved in Appendix~\ref{app:metrics}.

\begin{reptheorem}{thm:metrics}[Basic Version]

A clustering which maximizes the number of clusters also minimizes the number of county splits.
\end{reptheorem}

Does the court's metric give the clusterings with the most total clusters? In most cases, the answer is \textit{no}, i.e. there is usually a clustering which has more clusters than the ones deemed optimal by the courts that would lead to fewer county splits. To determine this, we expand our search tree to include clusterings which are slightly suboptimal at the $n$-county cluster level, but which may lead to better solutions at the statewide level, therefore ``relaxing'' the algorithm.

In the Senate with the current provisions, there are four ways to generate 29 clusters.
With the relaxed algorithm, we consider clusterings where the number of $n$-county clusters is allowed to be up to 3 fewer than the court-optimal solutions.\footnote{See Appendix~\ref{app:fuzz} for the exact alterations we make to the algorithm.}
By relaxing the criteria in the Senate, we find over 25000 clusterings with 29 clusters. 

\begin{figure}
    \centering
    \includegraphics{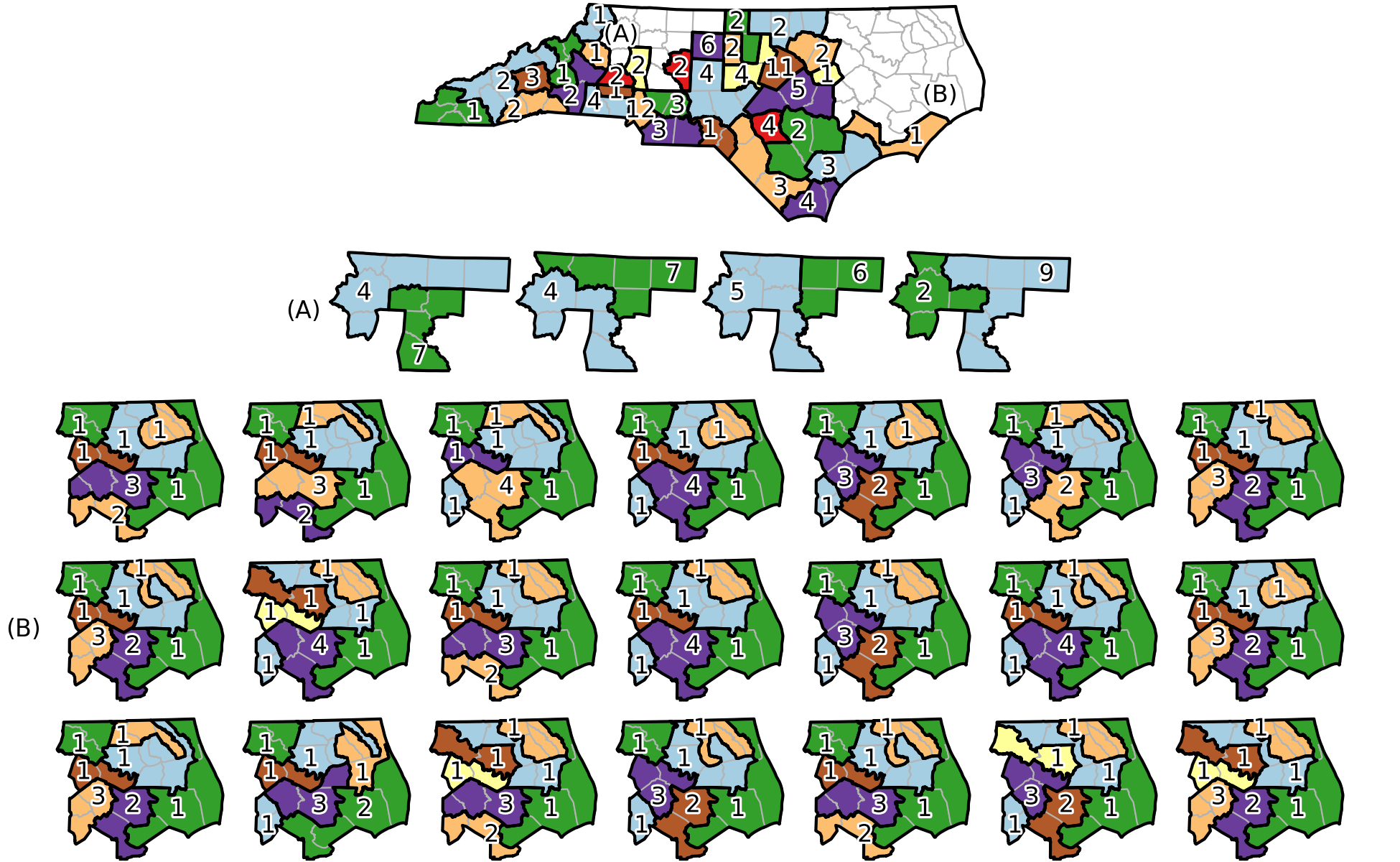}
    \caption{84 of the House clusterings that use 42 clusters under 2010 census data. These are not optimal under the court's metric, as they all contain two fewer 2-county clusters than the court-optimal clusterings, in return for more total clusters. There are at least 107 more House clusterings with 42 clusters than the ones shown; however, these other clusterings use fewer 1-county clusters.}
    \label{fig:House42}
\end{figure}

In the House there are two ways to generate 41 clusters with the current provisions; by relaxing the criteria as above, we find 191 clusterings with 42 clusters. In 84 of these House clusterings the number of singleton counties is the maximum possible; we display these 84 clusterings in Figure~\ref{fig:House42}. The alternative clusterings show that there are many more possible ways to cluster the counties.
Each of the 84 county clusterings contain two fewer 2-county clusters than the enacted plan, but more 3- and 4-county clusters, and the other 107 House clusterings we found use fewer 1-county clusters. This means they are not optimal according to the court's criteria. However, they admit one fewer county split than the enacted plans because they contain one more cluster. This shows that the court-optimal House clusterings do not maximize the total number of clusters.

In addition to the 2010 census data, the relaxed algorithm was run on all annual population estimates from 2010 to 2018. The results are summarized in Table~\ref{tab:algcompare}, where they are compared to the optimal clusterings according to the metric defined by the courts. In most cases, the relaxed algorithm finds clusterings that lead to fewer county splits than the court-defined optimal clusterings. It also usually finds vastly more possible clusterings.

\begin{table}[htb!]
    \centering
    \begin{tabular}{|c|r|r|r|r||r|r|r|r|}
        \cline{2-9}
        \multicolumn{1}{c|}{} & \multicolumn{4}{c||}{Senate} & \multicolumn{4}{c|}{House} \\
        \cline{2-9}
        \multicolumn{1}{c|}{} & \multicolumn{2}{c|}{Original alg.} & \multicolumn{2}{c||}{Relaxed alg.} & \multicolumn{2}{c|}{Original alg.} & \multicolumn{2}{c|}{Relaxed alg.} \\
        \hline
        Year & clusters & sol'ns & clusters & sol'ns & clusters & sol'ns & clusters & sol'ns \\
        \hline
        2010 Census & 29 & 4 & 29 & 25485 & 41 & 2 & 42 & 191 \\
        2010 Estimate & 25 & 4 & 30 & 1952 & 41 & 2 & 41 & 12220 \\
        2011 Estimate & 27 & 6 & 30 & 396 & 40 & 6 & 41 & 4796 \\
        2012 Estimate & 27 & 4 & 29 & 38850 & 37 & 3 & 41 & 2780 \\
        2013 Estimate & 27 & 2 & 30 & 328 & 39 & 12 & 40 & 552 \\
        2014 Estimate & 27 & 24 & 29 & 52189 & 39 & 6 & 40 & 256 \\
        2015 Estimate & 27 & 15 & 29 & 31070 & 40 & 16 & 40 & 3360 \\
        2016 Estimate & 28 & 12 & 29 & 9276 & 41 & 2 & 41 & 5778 \\
        2017 Estimate & 29 & 3 & 30 & 500 & 41 & 108 & 41 & 47360 \\
        2018 Estimate & 28 & 6 & 29 & 12224 & 39 & 6 & 41 & 1 \\
        \hline
    \end{tabular}
    \caption{Comparison of results from different criteria on the Whole County Provision. The relaxed algorithm seeks to maximize the number of clusters (therefore minimizing the number of county splits per Theorem~\ref{thm:metrics}), whereas the original algorithm implements the definition of optimality given by the courts. The relaxed algorithm does not guarantee to find all optimal solutions, nor guarantee the solutions it finds maximize the number of clusters; the original algorithm does guarantee that it finds all court-optimal solutions. The Senate and House clusterings found using the relaxed algorithm did not use the same parameters; see Appendix~\ref{app:fuzz} for further details.}
    \label{tab:algcompare}
\end{table}

The relaxed algorithm does not exhaustively search the space, and therefore does not guarantee the clusterings found lead to the highest number of clusters and therefore the fewest number of county splits. Significantly more computational power and algorithmic improvements would be necessary to provably find all clusterings which maximize the total number of clusters. The amount of time it takes to run the algorithm grows exponentially with the degree to which the search space is expanded. Nevertheless, a sufficiently relaxed algorithm would find all clusterings which maximize the total number of clusters. See Appendix~\ref{app:fuzz} for the exact alterations we make to the algorithm, further discussion of the extent to which the algorithm must be relaxed to find all optimal clusterings under this modified metric, and the computational feasibility of such relaxation.

Given the computational difficulty of finding the optimal clusterings under the alternate interpretation, it is not clear as a matter of public policy if it is better to have fewer split counties without an guarantee of optimality, or more split counties with a guarantee of optimality within the more narrow optimization procedure given by the court.  Furthermore, we have found that the number of possible cluster tends to increase dramatically under the expanded sense of optimal clusterings, begging the question of which to choose.

\subsection{Minimizing traversals}

In addition to the whole county provision, map makers for the North Carolina legislature must minimize number of \textit{county traversals}\cite{lakeStephensonBartlett2002,LegislatorGuideNorth2011}.  A district is said to traverse a boundary between two counties when there is a path within the district across the boundary.  In minimizing county traversals, one must not have more than one connected component of a district within a county.

Like minimizing the number of county splits, minimizing the overall number of county traversals is equivalent to maximizing the total number of clusters. We state and prove a more precise formulation of the following theorem in Appendix~\ref{app:metrics}.

\begin{repcorollary}{cor:traversals}[Basic Version]

A clustering which maximizes the number of clusters also minimizes the number of county traversals.
\end{repcorollary}

In short, all of the above work we have done to minimize the number of county splits is fully consistent with minimizing the total number of county traversals.

\section{Discussion}
\label{sec:Discussion}

We have developed an algorithm to cluster counties in a way that is consistent with court-adopted criteria. This algorithm has shown that the 2017 clusterings enacted by the North Carolina legislature are optimal, however it has also revealed that these clusterings are non-unique. We have developed and made our code available so that the clusterings adopted in the 2020 census cycle may be independently validated and compared with other possible choices \cite{Code}. We have also predicted what the 2020 county clusterings may look like based on projected census data.
We have investigated the stability of the clusterings and have found that even modest population changes can lead to significantly different possible county clusters. This suggests that clusterings will be sensitive to \textit{when} census data is collected and shows that the optimal clustering is not a fixed object.
Finally, we have considered an alternative criteria to adhere to the Whole County Provision, and have found that the other criteria generally leads to fewer split counties and significantly more possible county clusterings.
We have not, in this work, predicted how the choice of county clusterings, or county clustering criteria, may effect the partisan tilt in redistricting plans. We hope to consider this in future work.

\bibliography{biblio}

\begin{thebibliography}{FHIT18}

\bibitem[Ann19]{AnnualEstimatesResident2019}
Annual {{Estimates}} of the {{Resident Population}}: {{April}} 1, 2010 to
  {{July}} 1, 2018, May 2019.

\bibitem[Cod]{Code}
County clustering code.
\newblock https://git.math.duke.edu/gitlab/gjh/countycluster.git.

\bibitem[Cou]{CountyStatePopulation}
County/{{State Population Projections}}.
\newblock https://www.osbm.nc.gov/demog/county-projections.

\bibitem[Dic15]{DicksonRucho2015}
Dickson v. {{Rucho}}, December 2015.

\bibitem[FHIT18]{fifieldNewAutomatedRedistricting}
Benjamin Fifield, Michael Higgins, Kosuke Imai, and Alexander Tarr.
\newblock A {{New Automated Redistricting Simulator Using Markov Chain Monte
  Carlo}}.
\newblock May 2018.

\bibitem[Lak02]{lakeStephensonBartlett2002}
I.~Lake.
\newblock Stephenson v. {{Bartlett}}, April 2002.

\bibitem[Lak03]{lakeStephensonBartlettII2003}
I.~Lake.
\newblock Stephenson v. {{Bartlett II}}, July 2003.

\bibitem[Leg11]{LegislatorGuideNorth2011}
Legislator's {{Guide}} to {{North Carolina Legislative}} and {{Congressional
  Redistricting}}, March 2011.

\bibitem[Mei07]{meilaComparingClusteringsInformation2007}
Marina Meil{\u a}.
\newblock Comparing clusterings\textemdash{}an information based distance.
\newblock {\em Journal of Multivariate Analysis}, 98(5):873--895, May 2007.

\bibitem[Nor]{NorthCarolinaConstitution}
North {{Carolina Constitution}}.

\bibitem[Per17]{persilySPECIALMASTERRECOMMENDED2017}
Nathaniel Persily.
\newblock {{Special Master}}'{{s Recommended Plan and Report}}.
\newblock Technical report, December 2017.

\end{thebibliography}
\bibliographystyle{alpha}

\newpage

\begin{appendices}
\section{Mathematical and Algorithmic Details}
\subsection{Definitions and Notation}
\label{app:defs}

The \textit{county graph} $G$ is a graph where each vertex represents a county, and there is an edge between two vertices if the corresponding counties are adjacent. In North Carolina, the ruling is that counties \textit{can} be adjacent by water, but \textit{cannot} be adjacent by a single point (i.e. catty-cornered, or rook, rather than queen, adjacency). If $S$ is a set of counties, $pop(S)$ is the sum of the populations of $S$'s counties and $|S|$ is the number of counties in $S$. $pop(G)$ is thus the population of the entire state, and $|G|=100$ for North Carolina.

$G$ is to be divided into $D$ \textit{districts} such that the population of each district is within some fraction $\varepsilon$ of the ideal district population, given by $1/D$th of the total state population. In North Carolina, $\varepsilon=0.05$, i.e. up to 5\% deviation is acceptable, and $D=50$ for the state Senate and $D=120$ for the state House.

A \textit{county clustering} is a partition of $G$ into contiguous sets of counties with information about how many districts these sets should contain, called \textit{county clusters}. In other words, a county cluster is a tuple $(\text{set of counties}, \text{\# of districts})$, and a clustering is a set of clusters. A cluster is called \textit{valid} (or is said to have valid population) if its population divided by the number of districts it contains is within the error tolerance. In other words, if $G$ is to have $D$ districts, a cluster $(C,d)$ is valid if

\begin{align*}
\ceil*{(1-\varepsilon)\frac{pop(G)}{D}}d \le pop(C) \le \floor*{(1+\varepsilon)\frac{pop(G)}{D}}d\,,
\end{align*}
where $d$ is the number of districts in the cluster and $C$ is the set of counties in the cluster.
Note that in general the county set $C$ may have multiple choices of $d$ which satisfy these inequalities, especially if $C$ has very large population. A clustering is valid if all of its clusters $(C_1, d_1), (C_2, d_2), \dots$ are valid and

\begin{align*}
C_i \cap C_j &= \varnothing, \quad \forall i\neq j, \\
\sum_i d_i &= D\,, \quad \text{and}\\
\bigcup_i C_i &= G.
\end{align*}
As we have stated in Definition~\ref{def:optClust}, the courts have defined a clustering $A$ to be \textit{preferred} over another $B$ if $A$ has more 1-county clusters than $B$, or, if they have the same number, if $A$ has more 2-county clusters than $B$, and so on. In other words, if $A_n$ is the number of $n$-county clusters of $A$ and similarly for $B_n$, $A$ is preferred over $B$ if there is a $k$ such that $A_k>B_k$ and $A_j=B_j$ for all $j<k$. An \textit{optimal} or \textit{legal} clustering is one that is either preferred over or ties with all other clusterings.

\subsection{When Can Counties Be Clustered?}
\label{app:decision}

We now continue the discussion started in Section~\ref{sec:Algorithm}. We give a more precise version of Theorem~\ref{thm:decision} from that section as well as its proof.

\begin{theorem}[Enlarged Version]
\label{thm:decision}
Let $G$ be the county graph and $S$ be an induced subgraph of $G$. If $G$ is to contain a total of $D$ districts, $S$ has a valid clustering with a total of $d\le D$ districts if and only if both of the following are true:

\begin{itemize}
    \item All connected components $S_k$ of $S$ have valid population, i.e. for all $S_k$ there exists an integer $d_k$ such that
    \begin{align*}
     \ceil*{(1-\varepsilon)\frac{pop(G)}{D}}d_k \le pop(S_k) \le \floor*{(1+\varepsilon)\frac{pop(G)}{D}}d_k\,. 
    \end{align*}
    \item $d$ lies between the sum of the minimum number of districts each connected component $S_k$ of $S$ can contain and the related maximum, i.e.
    \begin{align*}
     \sum_k \ceil*{\frac{pop(S_k)}{\floor*{(1+\varepsilon)\frac{pop(G)}{D}}}} \le d \le \sum_k     \floor*{\frac{pop(S_k)}{\ceil*{(1-\varepsilon)\frac{pop(G)}{D}}}}\,.
    \end{align*}
\end{itemize}

\begin{proof}
Let $\{c_1, c_2, \dots\}$ be the set of connected components of $S$.

First, given the two statements of the theorem, we must prove that $S$ can be clustered into $d$ districts. We can rewrite the first statement of the theorem by noting that $d_k$ satisfies

\begin{align*}
\frac{pop(S_k)}{\floor*{(1+\varepsilon)\frac{pop(G)}{D}}} \le d_k \le \frac{pop(S_k)}{\ceil*{(1-\varepsilon)\frac{pop(G)}{D}}},   
\end{align*}
and since $d_k$ is integral, this is equivalently written
\begin{equation}
    \label{eqn:singlecc}
    \ceil*{\frac{pop(S_k)}{\floor*{(1+\varepsilon)\frac{pop(G)}{D}}}} \le d_k \le \floor*{\frac{pop(S_k)}{\ceil*{(1-\varepsilon)\frac{pop(G)}{D}}}}
\end{equation}
It is then easy to see that given the second statement in the theorem, there is some choice of $\{d_k^*\}$ which satisfy (\ref{eqn:singlecc}) for all $k$ and where $\sum_k d_k^*=d$. Then $\{(S_k, d_k^*)\}$ is a valid clustering of $S$ using $d$ districts.

Now for the reverse direction: given a valid clustering of $S$ using a total of $d$ districts, we must prove the two statements of the theorem. Let this valid clustering be the set of clusters $L=\{(C_1, d_1), (C_2, d_2), \dots \}$. Then we can partition $L$ into $\{L_k\}$, where $L_k=\{(C_i,d_i)\in L | C_i \subseteq S_k\}$. Then because the clustering $L$ is valid, for all $i$,

\begin{align*}
  \ceil*{(1-\varepsilon)\frac{pop(G)}{D}}d_i \le pop(C_i) \le \floor*{(1+\varepsilon)\frac{pop(G)}{D}}d_i\,.
\end{align*}
Then summing across connected components,
\begin{align*}
  \sum_{(C_i,d_i)\in L_k}\ceil*{(1-\varepsilon)\frac{pop(G)}{D}}d_i \le \sum_{(C_i,d_i)\in L_k}pop(C_i) \le \sum_{(C_i,d_i)\in L_k}\floor*{(1+\varepsilon)\frac{pop(G)}{D}}d_i 
\end{align*}
which implies
\begin{equation}
\label{eqn:cl-to-cc}
    \ceil*{(1-\varepsilon)\frac{pop(G)}{D}}\sum_{(C_i,d_i)\in L_k}d_i \le pop(S_k) \le \floor*{(1+\varepsilon)\frac{pop(G)}{D}}\sum_{(C_i,d_i)\in L_k}d_i
\end{equation}
so $(S_k, \sum_{(C_i, d_i)\in L_k} d_i)$ has a valid population for all $S_k$, which is the first condition in the theorem.

Now solve (\ref{eqn:cl-to-cc}) for $\sum_{(C_i, d_i)\in L_k} d_i$ and add in floor and ceiling symbols, as we did to find (\ref{eqn:singlecc}). The result is

\begin{align*}
  \ceil*{\frac{pop(S_k)}{\floor*{(1+\varepsilon)\frac{pop(G)}{D}}}} \le \sum_{(C_i, d_i)\in L_k} d_i \le \floor*{\frac{pop(S_k)}{\ceil*{(1-\varepsilon)\frac{pop(G)}{D}}}}\,.
\end{align*}
Then summing over all $k$,

\begin{align*}
  \sum_k \ceil*{\frac{pop(S_k)}{\floor*{(1+\varepsilon)\frac{pop(G)}{D}}}} \le \sum_k \sum_{(C_i, d_i)\in L_k} d_i \le \sum_k \floor*{\frac{pop(S_k)}{\ceil*{(1-\varepsilon)\frac{pop(G)}{D}}}}
\end{align*}
which implies
\begin{align*}
  \sum_k \ceil*{\frac{pop(S_k)}{\floor*{(1+\varepsilon)\frac{pop(G)}{D}}}} \le d \le \sum_k \floor*{\frac{pop(S_k)}{\ceil*{(1-\varepsilon)\frac{pop(G)}{D}}}}
\end{align*}
and the theorem is complete.

\end{proof}
\end{theorem}

Using this, we can quickly check each branch of the tree to see if the proposed clusters could possibly lead to a solution down the line.

\subsection{Optimizing the Search Tree}
\label{app:optimize}

In addition to the optimization given by the previous section, which effectively removes all branches which do not give solutions, two more major optimizations can be made to skip branches that give solutions that are known to not be optimal.

Let $S$ be the set of counties remaining at some branch of the tree, and let $V=\{v_1, v_2, \dots v_J\}$ be the set of all valid $n$-county clusters with counties contained in $S$. Explicitly, $v_i\in V$ is a cluster made up of $n$ counties, and the counties of $v_j$ may be shared with the counties of $v_i$; i.e. the county clusters of this set are not pairwise disjoint. Furthermore, the counties in the elements of $V$ do not cover $S$ in the general case, i.e. there may be some county in $S$ for which is not contained in a valid $n$-county cluster found in $S$.

Suppose we are finding the largest set of disjoint $n$-county clusters $S$ can contain, for some subgraph $S\subseteq G$, and we have already assigned as many 1-, 2-, ..., $(n-1)$-county clusters as possible. Let $\{S_k\}$ be the set of connected components of $S$. Then for all $S_k$ where

\begin{align*}
  n < |S_k| < 2n\,,
\end{align*}
we can remove from $V$ all $v_j$ where $v_j\subset S_k$. This is because if any $n$-county cluster was assigned on $S_k$, there would be less than $n$ left-over counties, but we already know we cannot form any clusters of less than $n$ counties.

Let $V'\subseteq V$ be the set of $n$-clusters that have not been pruned by the above optimization. Construct a new graph whose vertex set is counties in any cluster of $V'$; counties $c_i$ and $c_j$ share an edge if there is a $v_k\in V'$ such that $c_i,c_j\in v_k$. Let $\{V'_k\}$ be the connected components of this new graph. Then if we know that there is a collection made up of $T$ $n$-county clusters and the remaining unassigned counties, $S$, in the search tree is at depth $t<T$, we can skip the branch if we are unable to reach depth $T$ based on $V'$. To determine this, we note that each connected component of $V'$, $V'_k$ may be broken up into, at most, $\floor*{|V'_k|/n}$ $n$-county clusters. Therefore we can skip the branch if
\begin{align*}
  \sum_k \floor*{\frac{\left|V'_k\right|}{n}} < T-t\,, 
\end{align*}
as it will be impossible to find $T-t$ more $n$-county clusters.

Another way to look at this bound is that we first connect each county to other counties that could be part of the same $n$-county cluster. Connected components on this new graph represent disjoint sets of counties that don't interact at the $n$-county cluster level. Then, we skip any branch where the sum of the maximum number of $n$-county clusters of each connected component does not make up the difference between $T$ and $t$, so any leaf under that branch is necessarily a worse solution than we have already found.

\subsection{Complete Algorithm}
\label{app:algorithm}

The complete algorithm we use is as follows. First, we define Algorithm~\ref{alg:dfs}, which uses depth-first search and the optimizations described in the previous subsections. This performs the $(*)$ operation noted in Algorithm~\ref{alg:highlevel}. Then we have the iterative procedure defined in Algorithm~\ref{alg:complete} which outputs the set of all optimal clusterings, which is a fleshed out version of Algorithm~\ref{alg:highlevel}.

\begin{algorithm}[htb!]
    \caption{Recursive depth-first search to find largest valid $n$-county cluster sets}
    \label{alg:dfs}
    \KwIn{(always the same) Populations, adjacencies, minimum and maximum district population.}
    \KwIn{Set of counties $S$, number of districts $d$, cluster size $n$.}
    \KwIn{(for recursive calls only) Set of valid clusters $V$, deepest point reached so far $T$ (i.e. the number of $n$-county clusters we have found).}
    \KwOut{Set of all largest sets of $n$-county clusters with valid district numbers.}
    \If{not called recursively}{
        Let $V$ be the set of all valid $n$-county clusters on $S$ (these can be found simply by enumerating contiguous $n$-county subsets of $S$ and checking for validity)\;
        Let $T=0$\;
    }
    \If{the check given by Theorem~\ref{thm:decision} fails or the checks given in Appendix~\ref{app:optimize} fails}{
        \Return{$\varnothing$}\;
    }
    \tcp{Note that the checks in Appendix~\ref{app:optimize} may mutate $V$.}
    Let $O=\{\varnothing\}$\;
    \ForEach{$v_i\in V$}{
        Let $d_{min}=\ceil*{pop(v_i) / \text{max district population}}$\;
        Let $d_{max}=\floor*{pop(v_i) / \text{min district population}}$\;
        \ForEach{$d'\in [d_{min}, d_{max}]$}{
            Let $R$ be the output of Algorithm~\ref{alg:dfs} given inputs $S\gets S \setminus v_i$, $d\gets d-d'$, $n\gets n$, $V\gets \{v_j\in V | j>i, v_j\cap v_i=\varnothing\}$, $T\gets T$\;
            Let $R'=\{r\cup \{(v_i, d')\} | r\in R\}$\;
            \uIf{$|r|>T$ for arbitrary $r\in R$}{
                Set $O\gets R$\;
                Set $T\gets |r|$\;
            }\ElseIf{$|r|=T$ for arbitrary $r\in R$}{
                Set $O\gets O\cup R$\;
            }
        }
    }
    \Return{$O$}\;
\end{algorithm}

\begin{algorithm}[htb!]
    \caption{Complete algorithm}
    \label{alg:complete}
    \KwIn{Populations, adjacencies as a graph $G$, number of districts $D$.}
    \KwOut{All optimal clusterings.}
    Let $S$ be the vertex set of $G$, i.e. the set of all counties\;
    Let $O_0=\{\varnothing\}$\;
    Let $n=1$\;
    \While{$\bigcup_{(c,d)\in o} c \ne S$ for arbitrary $o\in O_{n-1}$}{
        Let $O_n=O_{n-1}$\;
        \ForEach{$o\in O_{n-1}$}{
            Let $d'=\sum_{(c,d)\in o} d$\;
            Let $S'=\bigcup_{(c,d)\in o} c$\;
            Let $R$ be the output of Algorithm~\ref{alg:dfs} given inputs $S\gets S\setminus S'$, $D\gets D-d'$, $n\gets n$\;
            Let $R'=\{r \cup o | r \in R\}$\;
            \uIf{$|r| > |o|$ for arbitrary $r\in R'$, $o\in O_n$}{
                Set $O\gets R'$\;
            }\ElseIf{$|r| = |o|$ for arbitrary $r\in R'$, $o\in O_n$}{
                Set $O\gets O\cup R'$\;
            }
        }
        Increment $n$\;
    }
    \Return{$O_{n-1}$}\;
\end{algorithm}

\section{Clustering Distances}
\label{app:stability}

Consider clustering $A$ made of clusters $A_1, A_2, \dots$ and clustering $B$ made of clusters $B_1, B_2, \dots$, and there are a total of $n$ counties. The \textit{different clusters} is defined as

\begin{align*}
  DC(A,B) = 100\%\cdot \left(1-\frac{|A\cap B|}{\frac{1}{2}(|A| + |B|)}\right)
\end{align*}

The \textit{variation of information}, as defined in \cite{meilaComparingClusteringsInformation2007} and used for similar purposes in \cite{fifieldNewAutomatedRedistricting}, is

\begin{align*}
    VI(A,B) &= H(A|B) + H(B|A) \\
    &= -\sum_{i,j} \frac{|A_i \cap B_j|}{n} \log\frac{|A_i \cap B_j|^2}{|A_i||B_j|}\,,
\end{align*}
and the log is base 2 so the units are \textit{bits/county}. The summand is defined to be 0 if $A_i$ and $B_j$ are disjoint. Note that $VI$ and $DC$ are both symmetrical in their arguments, and they are zero if $A=B$. $VI$ is a true metric \cite{meilaComparingClusteringsInformation2007}, but $DC$ does not satisfy the triangle inequality.

Consider year $X$ where the populations of counties are $X_1, X_2, \dots, X_n$ and year $Y$ where the respective populations are $Y_1, Y_2, \dots, Y_n$. The \textit{average population change} is defined as

\begin{align*}
  APC(X,Y) = \frac{100\%}{n}\sum_{i} \frac{|X_i - Y_i|}{\frac{1}{2}(X_i + Y_i)}\,. 
\end{align*}

The reason that the denominator of the summand is the average of the two populations rather than the earlier population (as population changes are typically expressed) is to make the function symmetrical.

\section{Other Optimality Metric}
\label{app:metrics}
We now return to Theorem~\ref{thm:metrics} which was previously discussed in Section~\ref{sec:metrics}. We give and prove a more precise version of this theorem
\subsection{Proof of Theorem~\ref{thm:metrics} and Corollary~\ref{cor:traversals}}

Suppose in a districting $\xi$, county $c_i$ contains parts or all of $m_i$ districts. Then the number of \textit{county splits} $D$ has is defined to be $f(\xi)=\sum_i (m_i-1)$.

We introduce the notion of a \textit{minimal cluster}. A cluster is minimal if it is valid and cannot be split into two smaller valid clusters. Note that the clustering which maximizes the total number of clusters contains only minimal clusters. Now we have the theorem:

\begin{theorem}[Enlarged Version]
\label{thm:metrics}
The clustering which maximizes the number of clusters, when districted optimally, also minimizes the number of county splits, with the exception of rare circumstances which impact the optimal districting.

\begin{proof}

We proceed by ``growing'' a cluster to show that minimal clusters with $d$ districts, when districted optimally, usually contribute $d-1$ county splits. Consider a cluster $(C, d)$, where $C=\{c_1, c_2, \dots c_n\}$, a set of counties and $d$ is a number of districts. The set $C$ induces a subgraph of the county graph $G$. On this subgraph, draw a rooted spanning tree $T$. Order the vertices such that each vertex comes before its parent in the tree, and the root node is last. We will use this spanning tree and ordering to construct a districting with few county splits.

\vspace{1em}

Consider the vertices in the order decided. We assign districts to a vertex according to three cases: the vertex is a non-root leaf, the vertex is a non-root non-leaf, and the vertex is the root.

Consider a leaf of the spanning tree $c_i$. Let it contain some number of whole districts and exactly 1 ``partial'' district, which is shared with other counties. Draw the whole districts such that the remainder of the county is connected and contains at least part of the border with $c_i$'s parent in the spanning tree.

Consider a county $c_j$ which is not a leaf and not the root. If the partial districts of two children of $c_j$ combine to less than a valid district and those children are ``next to'' one another, extend their partial districts inside $c_j$, connecting them. Consider this combination as one ``child'' from this point forward. The requirement that the two children are ``next to'' each other means that, upon connecting them, $c_j$ is not broken into multiple disconnected components. Continue connecting children until as many as possible are connected. When we connect two children in this manner, we call it a \textit{good combine}.

Next use $c_j$'s population to complete as many partial districts as possible. After doing this, there are two cases: either all partial districts were completed and $c_j$ has population left over, or there are some number of uncompleted children and $c_j$'s remaining population is less than the population of one district. In the former case, draw as many whole districts on $c_j$ as possible, leaving the remainder of $c_j$ connected and bordering its parent. The latter case has two subcases: there is either one child left, or multiple children left. If only one child, combine this child with the rest $c_j$, with the new partial district bordering $c_j$'s parent.

If more than one child is left, it must be the case that partial districts which are next to each other sum to more than one district even without any of $c_j$'s population. Call two adjacent children $x$ and $y$. Use a small amount of $c_j$ to add to $y$'s partial district $\xi_y$ and bridge over to $x$. Then, reapportion part of $x$'s partial district $\xi_x$ to now be part of the $\xi_y$, leaving $\xi_y$ with valid population and $\xi_x$ with less population. Note that $\xi_y$ now stretches across three counties: $x$, $c_j$, and $y$. On the other hand, $\xi_x$ contains part of $x$ and borders $c_j$.

What we have done is combined children $x$ and $y$, but added one county split. We will call this a \textit{bad combine}. After doing a bad combine, it may be possible to do more good combines. Continue combining children, first doing as many good combines as possible, then one bad combine, then good combines, etc. until only one child remains. Combine this child with the rest of $c_j$, whose partial district will end up bordering $c_j$'s parent.

Repeat this process until just one county, the root county, remains. For this county, as before, connect children, complete as many partial districts as possible, then connect children, and so on. As all districts need only to have valid population in some range, rather than a specific value, there may be slightly too much or too little population in the root county to complete the districting. In these cases, return to lower layers of the tree and tweak boundaries slightly while staying with good population and not changing which districts are part of which counties, until the districting can be completed.

\vspace{1em}

We will now compute the number of county split this process resulted in. Each edge of the spanning tree represents exactly one time that a district was part of multiple counties, with the exception of edges that were used in bad combines, which contribute one more for each bad combine. Hence, summing the number of districts each county contains ($m_i$) over all counties in the cluster gives the number of districts, plus the number of edges in the spanning tree, plus the number of bad combines. Thus,

\begin{align*}
    \text{\# of county splits in cluster} &= \sum_{c_i\in C} (m_i-1) \\
    &= \left(\sum_{c_i\in C} m_i\right) - |C| \\
    &= d + (|C| - 1) + (\text{\# of bad combines}) - |C| \\
    &= d - 1 + (\text{\# of bad combines})\,.
\end{align*}

Given a districting which minimizes county splits, one can construct a spanning tree, vertex ordering, and sequence of combines such that this process results in the given districting, so performing this process for all spanning trees, orderings, and sequences of combines is guaranteed to minimize county splits. Although it is not difficult to construct examples where bad combines are necessary in all cases,\footnote{The smallest example contains four counties: one in the middle and three surrounding it, which receives two districts total. None of the three outer counties touch each other, only the inner county. Then if the outer counties have populations 0.9, 0.9, and 0.2 of a district, with the inner county having negligible population, one bad edge combine is required. \\ For a bad edge combine to be required, it is necessary that (1) the cluster receives multiple districts, and (2) for \textit{all} spanning trees on the cluster, there is a vertex with ``low population'' and degree at least 3.} we think it is rare to require any in most real-world scenarios. Hence, if a cluster contains $d$ districts, it usually can be districted using $d-1$ county splits.

On the other hand, $d-1$ county splits is a lower bound for minimal clusters containing $d$ districts. For any districting, we can consider the subgraph generated by the districting, where two counties are adjacent in the subgraph if they are adjacent in $G$ and a district contains parts of both of them. Because the cluster is minimal, this graph must be connected (otherwise the cluster could be split via the connected components into multiple valid clusters). Hence, it contains at least $E\ge |C|-1$ edges. By a similar calculation to above, the number of county splits is at least $d+E-|C|$, which is at least $d-1$.

Now the total number of county splits in all clusters is
\begin{align*}
  \sum_{\text{counties } c_i} (m_i-1) \ge \sum_{\text{clusters}(C_j, d_j)} (d_j-1)=\left(\sum_{\text{clusters }(C_j, d_j)} d_j\right) - \text{\# of clusters}\,. 
\end{align*}

The first term in the right-hand expression just the total number of districts and is constant. We suspect the inequality is almost always tight in real life; if this holds, then maximizing the number of clusters minimizes the number of county splits.

\end{proof}
\end{theorem}

\begin{corollary}[Enlarged Version]
\label{cor:traversals}
The clustering which maximizes the number of clusters, when districted optimally, also minimizes the number of traversals, with the exception of rare circumstances which impact the optimal districting.

\begin{proof}
Each edge in the spanning tree in the proof of Theorem~\ref{thm:metrics} represents a district traversing a county-to-county boundary (i.e. one traversal).  The only additional requirement when optimally redistricting is that when partial districts in the children are combined with sections of the parent, the district within the parent must form a connected component.  The only exception to an edge adding only one traversal are the \textit{bad combine}'s which introduce two traversals.  The number of edges in a spanning tree is one less the number of vertices
\begin{align*}
\text{\# of traversals in cluster} = |C_j|-1 + (\text{\# of bad combines}),
\end{align*}
where $|C_j|$ is the number of counties in cluster $j$.
Therefore, the total number of traversals is 
\begin{align*}
\text{\# of traversals} \ge \sum_{\text{clusters}(C_j, d_j)} (|C_j| - 1) = |C| - \text{\# of clusters}\,.
\end{align*}

The first term in the right-hand expression just the total number of counties and is constant. As mentioned in the previous proof, we suspect bad traversals to be rare, making the  inequality almost always tight in real life; if this holds, then maximizing the number of clusters minimizes the number of traversals.

\end{proof}
\end{corollary}

\subsection{Fuzziness}
\label{app:fuzz}

The alterations to the algorithm are fairly straightforward. First, we change the metric used to bound branches, because simply comparing the number of clusters does not account for counties remaining. 
As an example, consider a partial clustering which uses 1 single-county cluster and no two-county clusters, compared to a partial clustering which used no single-county clusters but 2 two-county clusters. 
Which of these clusterings is preferred? 
Although the second clustering uses more clusters, it has fewer counties left to cluster so it may lead to fewer clusters in the long run. In fact, the first case has the potential to have one more three-county cluster than the second case. If the second case instead had 3 two-county clusters, or if it was known that the first case did not lead to one more three-county cluster, the second case would be better. With this in mind, the metric we now used is

\begin{align*}
  (n+1)(\text{\# of clusters}) + (\text{\# of unassigned counties})\,,   
\end{align*}
where $n$ is the number of counties in the clusters just assigned; in the previous example, $n=2$. Note that adding one cluster containing $n$ counties increases the first term by $n+1$ and decreases the second term by $n$, for a net increase of 1. Under this metric, the two cases described previously would be equal after $n=2$, though if $n$ increased without the first case ``catching up,'' the second case is better.

Next we introduce a ``fuzziness'' $f$. At each iteration, instead of taking all solutions which maximize this measure, we take all solutions that are at least $f$ smaller than the best solution under this measure. Almost no changes are needed to adapt the bounds and optimizations to this measure. The first optimization in Appendix~\ref{app:optimize} can still be done; although it is not guaranteed that all possible $n$-county clusters are taken at each step, at no future point in the algorithm will any other $n$-county clusters be created. The second optimization calculates the maximum number of clusters that could be added. Since each cluster increases the measure by 1, the calculation is exactly the same, though now we only skip a branch if we cannot make up the difference between $t$, the current depth, and $T-f$, the greatest depth reached minus $f$. The other alteration to the algorithm is in updating $O$ (in the notation of Algorithms~\ref{alg:dfs} and \ref{alg:complete}); at each step we now set $O$ to be all elements of $O$ and $R'$ which are at least as good as the best element of those sets minus $f$.

\subsubsection{Details of our analysis}

In our analysis, a fuzziness of 3 was used for the Senate and a fuzziness of 2 was used for the House. This is because the House has a much larger number of valid clusters, leading to much larger search trees than the Senate.

Additionally, for both the Senate and House except in the 2010 census House, an optimization was made which reduces the number of cases considered, which is as follows. Consider a case where there are two regions, A and B, that each have two options, (A1 and A2) and (B1 and B2). Only two clusterings are needed to reconstruct all four possibilities, namely the clustering with A1 and B1 and the one with A2 and B2 (or A1, B2 and A2, B1). For larger sets, such as those dealt with by the relaxed algorithm, a significant amount of time and memory is saved by considering only the reduced set of clusterings which contain enough information to reconstruct the whole set.

The complete set can be reconstructed by comparing each pair of clusterings and checking for situations like in the above paragraph describes. Additional solutions that were reconstructed must also be compared against all other solutions, and so on until no more additional solutions are found.

\subsubsection{How much fuzziness is required?}

Clearly, having a fuzziness of 100 (the number of counties in North Carolina) would suffice to provably find all maximum-cluster clusterings. However, the same result can be accomplished with much less fuzziness. First, consider the House using 2010 census data. It is known that there exists a clustering with 42 clusters. The legal precedent requires taking all 1-county clusters, of which there are 12. If a proposed partial clustering had no 2-county clusters, it could have at most $12+\floor*{88/3}=41$ clusters. In fact, at least two 2-county clusters must be taken in order to tie 42; at least 5 must be taken to get 43. The most possible 2-county clusters is 17 for the House, so a fuzziness of $17-5=12$ suffices to prove that 42 clusters is optimal.

In fact, this can be improved by noting that not all counties can be part of 3-county clusters, and some must be part of 4- or 5-county clusters. In particular, the westernmost county Cherokee must combine with at least 3 others, as do many of the eastern counties with low population. In fact, one can show that at least 14 counties must be part of a cluster with at least 4 counties. All this means a fuzziness of $17-10=7$ suffices to prove 42 clusters is optimal (43 clusters could be reached under these constraints, for example, by 12 1-county clusters, 10 2-county clusters, 18 3-county clusters, 2 4-county clusters, and a 6-county cluster).

It is harder to find good bounds for the Senate. In particular, the additional analysis in the previous paragraph is far more complicated as a result of an increased number of possibilities. Here is a simpler analysis. There is a unique 1-county cluster (Mecklenburg). In the case that no 2-county clusters are taken, at most 16 3-county clusters can be taken, resulting in at most 29 clusters (11 4-county and one 7-county in addition to Mecklenburg and the 16 3-county clusters), so a fuzziness of $13$ (based on the maximum number of 2-county clusters) suffices.

Reaching a fuzziness of 7 for the House or 13 for the Senate is out of reach without major alterations to the algorithm. Both the time and space complexity are empirically at least exponential in fuzziness, so it would take orders of magnitude longer and more memory to run the algorithm when compared to the analysis we completed using fuzziness 2 and 3. More research would also be required to develop a good general algorithm for bounding fuzziness in order to extend this algorithm to other population datasets.

\end{appendices}

\end{document}